\documentclass[RNAAS]{aastex62}

\received{2018 June 18}
\submitjournal{RNAAS}

\shortauthors{Lucey et al.}
\begin{document}

\title{A new quadruple-image gravitational lens in an edge-on disk galaxy at z=0.0956}

\correspondingauthor{John R.\ Lucey}
\email{john.lucey@durham.ac.uk}

\author[0000-0002-9748-961X]{John R.\ Lucey}
\affil{Centre for Extragalactic Astronomy, Durham University, Durham DH1 3LE, United Kingdom}

\author[0000-0001-5998-2297]{Russell J.\ Smith}
\affil{Centre for Extragalactic Astronomy, Durham University, Durham DH1 3LE, United Kingdom}

\author[0000-0002-5665-4172]{Paul L.\ Schechter}
\affil{MIT Kavli Institute 37-664G, 77 Massachusetts Avenue, Cambridge, MA, 02138-4307, USA}

\author[0000-0003-4772-528X]{Amanda S.\ Bosh}
\affil{MIT Department of Earth, Atmospheric, and Planetary Sciences, Cambridge, MA  02139, USA}

\author[0000-0002-1050-3539]{Stephen E.\ Levine}
\affil{Lowell Observatory, Flagstaff, AZ  86001, USA}


\section{Background and discovery}

\begin{figure}[t]
\plotone{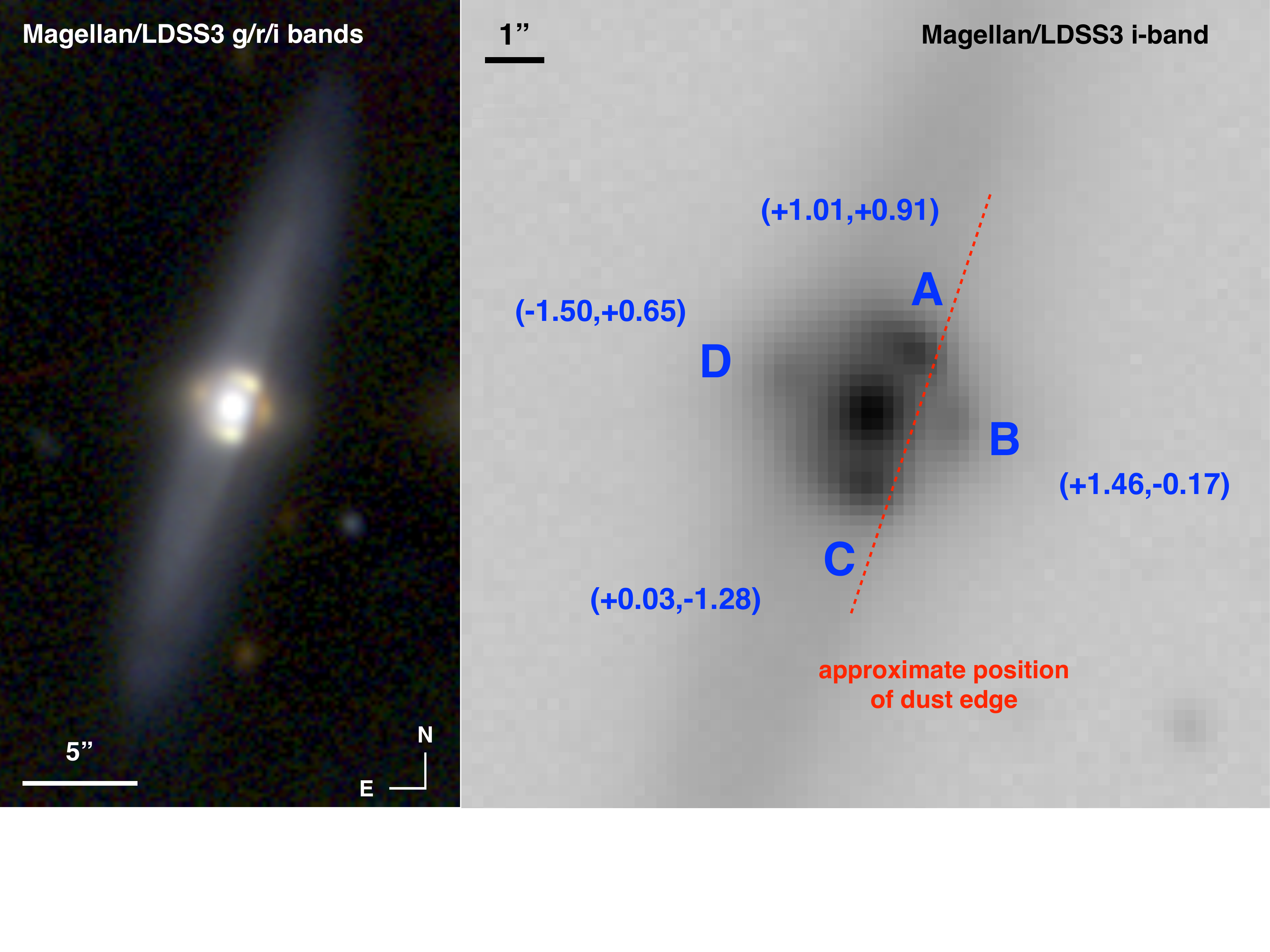}
\vskip -15mm
\caption{Left: Pseudo-true-colour image of 2MASXJ13170000--1405187 from Magellan $gri$ imaging. Right: Zoomed $i$-band image. Source positions are
in arcsec relative to the bulge centre, derived from the VHS K-band data.}
\end{figure}

Strong gravitational lensing is invaluable as a probe of the mass distribution within galaxies, providing insights into both the mass-to-light ratio of stellar populations and the
structure of dark-matter halos, especially when combined with stellar dynamical measurements \citetext{e.g. \citealt{2010ApJ...724..511A}}.
Most strong-lens samples are dominated by massive early type galaxies, which are favoured by several selection effects.
However spiral lenses, in particular those with edge-on disks, are keenly sought, since their rotation curves provide relatively 
direct kinematic information to constrain more complex lens models \citetext{e.g. \citealt{2011MNRAS.41.1601T}}.

In order to refine the input catalogue for the Taipan Galaxy Survey \citep{2017PASA...34...47D}
optical image cutouts from a variety of legacy surveys,
e.g. Pan-STARRS PS1 \citep{2016arXiv161205560C}, are being visually inspected by one of us (JRL).
From this work we have previously published the discovery of two
quadruple systems \citep{2018MNRAS.476..927L},
both of which have been confirmed spectroscopically as lensed quasars \citep{2018arXiv180307175R}.

Here we report the further serendipitous discovery of a
quadruply-lensed source at the core of a $z$\,=\,0.0956 edge-on disk galaxy, 2MASXJ13170000--1405187.
Features suggestive of a lensing configuration were initially noted in the inspection of PS1 data, and confirmed in 
$K$-band imaging from the VISTA Hemisphere Survey (VHS; \citealt{2013Msngr.154...35M}).
We subsequently obtained deeper optical observations with LDSS3 at the 6.5m Magellan Clay telescope
in $g$, $r$ and $i$ (360\,sec in each band), with FWHM\,$\approx$\,0\,\farcs6.
Extracts from these images are shown in Figure~1. 

\vskip 4mm

\section{Properties of the lens galaxy}

2MASXJ13170000--1405187 is an edge-on disk galaxy, with a redshift of $z$\,=\,0.0956
(6dFGS, \citealt{2009MNRAS.399..683}).
The galaxy spectrum is continuum-dominated, with the strong metal lines and weak Balmer absorption typical of old stellar populations.
There is weak emission at  [O\,{\sc ii}] 3727\,\AA\ and  [N\,{\sc ii}] 6584\,\AA.

The Magellan imaging shows that the galaxy has a bright central bulge with radius $\sim$1\arcsec\ (1.8\,kpc) and 
an extended disk with low surface brightness, traceable to a radius of $\sim$15\arcsec\ (27\,kpc). 
The $g$-band image reveals a prominent dust ring at a major-axis radius of $\sim$9\arcsec.
The apparent axial ratio suggests a disk inclination of $i$\,$\ga$\,80$^\circ$. 

\vskip 4mm

\section{Lensing configuration and preliminary modelling}

The four lensed images (A--D) are observed at radii 1.0--1.5\,arcsec from the bulge centre. 
The images are clearly extended, showing that the background source is a galaxy, not a quasar. 
All of the lensed features, as well as the galaxy nucleus and inner disk, would have contributed to the 6dFGS spectrum (fibre 6\,\farcs7 diameter), 
but no background source lines can be identified in the spectral range covered (3950--7550\,\AA).

Images B and D lie behind the dust ring identified in the $g$-band image, and are both subject to significant reddening. 
Images A and C extend into the dust ring at their western ends, which hampers deriving accurate positions to 
use in modelling the lens. 
To mitigate against this, we use positions measured from the VHS K-band image, derived by PSF-fitting following \cite{1993PASP..105.1342S}, 
and summarized in Figure~1.

We use {\sc lensmodel} \citep{2001astro.ph..2340K} to fit a simple lensing model to the VHS-derived positions, using Monte Carlo realisations to propagate positional errors of 0.25\,pixels.
The model is a singular isothermal ellipse with free 
centre, ellipticity, position angle, and normalisation. The best-fitting model has ellipticity 1--$b/a$\,=\,0.4$\pm$0.1 at position angle 19$\pm$2$^\circ$ (consistent with the disk orientation). 
The circularized Einstein radius is 1.44$\pm$0.05\arcsec\ (2.57$\pm$0.09\,kpc),
corresponding to a projected mass of 
(9.4$\pm$0.6)\,$\times$10$^{10}$\,$M_\odot$ 
(assuming the background source is much more distant than the lens).

The observed K-band luminosity within the Einstein radius is 24$\times$10$^{10}$\,$L_\odot$, which overestimates the galaxy contribution 
as it includes substantial flux from the lensed images.
Using the higher-resolution $i$-band image to estimate this contamination, 
our best estimate of the Einstein aperture luminosity of the lens galaxy itself is $L_K$\,=\,18$\times$10$^{10}$\,$L_\odot$.
The resulting mass-to-light ratio range of $\sim$0.5\,$(M/L_K)_\odot$ is consistent with 
intermediate-age solar-metallicity stellar populations, e.g. 4\,Gyr with Kroupa IMF, or 2.5\,Gyr with Salpeter IMF \citep{2005MNRAS.362..799M}.

\vskip 4mm

\section{Future prospects}

Future exploitation of 2MASXJ13170000--1405187 will require spectroscopy to establish the background source redshift, and to constrain better the stellar populations in the 
lensing galaxy. High-resolution imaging in the infra-red, to reduce the impact of the dust obscuration, will be essential for pixel-based lens modelling.
Adaptive optics observations should be possible, as there is a $r$\,$\approx$\,14 star located 30\arcsec\ south of the target. 

\acknowledgments

\noindent
\vskip -4mm
We thank all our co-workers who have contributed to the legacy image surveys used in this analysis.

\vspace{1mm}
\noindent
\facilities{PS1, ESO:VISTA, Magellan:Clay}

\end{document}